# Experimental and theoretical assessment of native oxide in the superconducting TaN

V. Quintanar-Zamora[1,2], M. Cedillo-Rosillo[1,2], O. Contreras-López[2], C. Corona-García[2], A. Reyes-Serrato[2], R. Ponce-Pérez[2], J. Guerrero-Sánchez[2], J. A. Díaz[2]

1. Centro de Investigación Científica y de Educación Superior de Ensenada, Baja California, Ensenada, México
2. Centro de Nanociencias y Nanotecnología, Universidad Nacional Autónoma de México, Ensenada, México

**Abstract**

In this manuscript, we show through an experimental-computational proof of concept the native oxide formation into superconducting TaN films. First, TaN was synthesized at an ultra-high vacuum system by reactive pulsed laser deposition and characterized *in situ* by X-ray photoelectron spectroscopy. The material was also characterized *ex situ* by X-ray diffraction, transmission electron microscopy, and the four-point probe method. It was detected that TaN contained considerable oxygen impurities (up to 26 %O) even though it was grown in an ultra-high vacuum chamber. Furthermore, the impurified TaN evidence a face-centered cubic crystalline structure only and exhibits superconductivity at 2.99 K. To understand the feasibility of the native oxide in TaN, we study the effect of incorporating different amounts of O atoms in TaN using ab-initio calculations. A thermodynamic stability analysis shows that a $TaO_xN_{1-x}$ model increases its stability as oxygen is added, demonstrating that oxygen may always be present in TaN, even when obtained at ultra-high vacuum conditions. All analyzed models exhibit metallic behavior. Charge density difference maps reveal that N and O atoms have a higher charge density redistribution than Ta atoms. The electron localization function maps and line profiles indicate that Ta-O and Ta-N bonds are mainly ionic. As expected, stronger ionic behavior is observed in the Ta-O bonds due to the electronegativity difference between O and N atoms. Recent evidence points to superconductivity in bulk TaO, confirming the asseverations of superconductivity in our samples. The results discussed here highlight the importance of considering native oxide when reporting superconductivity in TaN films since the TaO regions formed in the compound may be key to understanding the different critical temperatures reported in the literature.

Keywords: DFT, tantalum nitride, tantalum oxynitride, superconductivity, laser ablation, thin films, XPS.

## 1. Introduction

Superconducting materials have a wide range of applications, including superconducting devices. In this regard, superconducting tantalum nitride (TaN) has attracted attention due to its applications in the detection of low-energy photons in superconducting nanowire single-photon detectors (SNSPD) [1] and in the development of normal metal-insulator-superconductor (NIS) tunnel junctions [2].

Tantalum nitride is a transition metal nitride (TMN) that presents a wide variety of crystalline phases, for example, $\beta$-TaN$_{0.05}$, $\gamma$-Ta$_2$N, $\delta$-TaN$_{1-x}$, $\varepsilon$-TaN, Ta$_5$N$_6$, Ta$_4$N$_5$, Ta$_3$N$_5$, among others [3-6]. Within their different polymorphs, only $\delta$-TaN$_{1-x}$ (which from here we will call TaN) presents superconductivity at a critical temperature (T$_c$) of 10.8 K [7]. This compound crystallizes in a NaCl-type structure with a= 4.34 Å [8, 9]. Despite being metastable [10], several groups have reported its growth by reactive sputtering [4, 7, 11-13] and laser ablation [14].

Gerstenberg *et al.* synthesized for the first time superconducting TaN thin films with a T$_c$ of 4.84 K measured resistively [4]. Subsequently, Boiko & Popova obtained bulk superconducting TaN with a T$_c$ of 6.5 K measured by a "magnetic method" [15]. Afterward, other authors continued to produce superconducting TaN thin films within a wide range of T$_c$ from 0.5 K to 10.8 K [7, 11-14]. However, in these reports, the presence of native oxide has been overlooked. Still, the T$_c$ vary considerably, ranging between 0.5 K to ~11 K.

Experimental evidence points out that TaN thin films present oxygen impurities regardless of the synthesis method or the employed gas mixture [16]. This material is susceptible to oxygen contamination, even when grown at ultra-high vacuum and analyzed *in situ* [17]. On the other hand, oxynitrides can exist in wide compositional ranges [18]. Their physical properties are sensitive to composition [19]. Therefore, adjusting the nitrogen/oxygen ratio to obtain tantalum oxynitrides can lead to potentially attractive applications [20]. These materials have received attention due to their application as pigments [21] and their efficient use in water splitting [22].

Cristea *et al.* reported face-centered cubic (FCC) tantalum oxynitride ($TaO_xN_y$) thin films growth by DC reactive magnetron sputtering [23]. They determined the atomic composition of the samples by Rutherford backscattering spectrometry (RBS) and observed only the peaks corresponding to an FCC phase in X-ray diffraction (XRD) patterns. They also studied other tantalum oxynitrides that present corrosion resistance [24], antibacterial capacity, and photocatalytic activity [25]. However, the FCC tantalum oxynitride has yet to be described at the atomic scale because there are no simulations of the possible sub-stoichiometries in the cubic phase. So, we treat an FCC $TaO_xN_{1-x}$ $Fm\bar{3}m$ phase in compositions ranging between $0 \leq x \leq 1$.

First, we grow tantalum nitride thin films at ultra-high vacuum conditions and analyze their composition *in situ*. Samples possess a considerable amount of oxygen, but the FCC structure remains, and it depicts superconductivity at 2.99 K, a temperature within the well-known range reported in the literature for tantalum nitride. Since oxygen appears on the samples even at ultra-high vacuum conditions, we determine the stability of various FCC $TaO_xN_{1-x}$ sub-stoichiometries with different amounts of oxygen ($0 < x \leq 0.50$) by density functional theory (DFT) calculations and analyze the impact of the oxygen content on their structural and electronic properties. In each case, oxygen atoms take on nitrogen sites in the TaN crystal structure for a large growth range. We also considered the limit cases when x takes values of 0 or 1. Metallic characteristics emerge in each oxide concentration. Both Ta-N and Ta-O bonds are ionic, but Ta-O shows a stronger ionic character. The evidence of native oxygen in tantalum nitride films points to the importance of discussing the critical temperature in terms of the tantalum oxygen regions formed in the films.

## 2. Methods

### 2.1 Experimental procedure

The TaN thin film was synthesized by reactive pulsed laser deposition (PLD) in a modified RIBER LDM 32 ultra-high vacuum system with a base pressure of 5x10$^{-9}$ Torr. A 99.99% pure tantalum target from Kurt J. Lesker was employed to obtain the Ta atoms. A 6 ns and 266 nm pulse from a Surelite Continuum Nd:YAG solid-state laser was used to ablate the target. The laser has a fundamental wavelength of 1064 nm, a second harmonic generator of 533 nm, and a fourth harmonic generator of 266 nm. The working pressure of the chamber was set at 0.09 Torr of ultra-high purity $N_2$ gas (99.999%). Then, the tantalum nitride thin film was deposited on FCC MgO (100) substrate from Sigma-Aldrich to achieve epitaxial growth at 850 ºC for 220 min.

The chemical composition was analyzed *in situ* by X-ray photoelectron spectroscopy (XPS). All other characterizations were measured *ex situ*. The XPS spectra were registered with a SPECS PHOIBOS hemispherical analyzer, where the X-rays were produced with a monochromatic source of Al Kα (1486.6 eV). Firstly, a low-resolution spectrum was measured at pass energy of 100 eV and a data acquisition time of 0.1 s to identify the elemental components of the film. Next, the high-resolution spectra of the Ta 4f, N 1s, and O 1s peaks were measured at a pass energy of 10 eV and a data acquisition time of 0.5 s to quantify the previously identified components; the Ta $4p_{3/2}$ peak was also measured because it overlaps with N 1s. Then, background noise was subtracted by the Shirley method [26], and the peak area of each component was calculated separately from a curve fitting. In addition, the Ta chemical environment was identified from the Ta 4f peak. The NIST Standard Reference Database 20 [27] was used as a reference for peak identification. All the spectra were processed with the CasaXPS software [28].

The crystal structure was subsequently identified by XRD with a Panalytical X'pert Pro MRD diffractometer in the 2θ range from 30º to 80º using a glancing angle of 0.1°, a step of 0.02°, and a data acquisition time of 0.5 s. The X-rays were produced by a Cu Kα source (λ = 1.5406 Å). The sample reflections were indexed with the ICDD database. The data were processed with the HighScore Plus software [29]. In addition, the local crystal structure was explored with a JEOL JEM-2100F scanning transmission electron microscope (STEM), which has a Schottky-type field emission electron gun (FEG) with a 200 kV electron beam corresponding to a wavelength of around 0.0025 nm. The specimen was analyzed in conventional parallel-beam mode. Then, high-resolution micrographs of the cross-section of the thin film and its electron diffraction pattern were recorded. The sample was prepared with a JEOL JIB-4500 scanning electron microscopy focused ion beam (SEM-FIB) to reduce the film cross-section thickness to less than 100 nm by micro-milling with gallium ions. The micrographs and the diffraction pattern were processed with the Digital Micrograph software.

The resistance vs. temperature curve was measured with the four-point probe method in a Quantum Design DynaCool Physical Properties Measurement System (PPMS) to identify the superconducting $T_c$. The sample was cooled down from room temperature to 2 K and measured at a constant direct current of 0.05 mA.

**2.2 DFT calculations**

The atomic-scale description of the $TaO_xN_{1-x}$ was performed in the Vienna Ab-initio Simulation Package (VASP) [30-35] code. The electron-ion interactions were treated employing the Projector Augmented Wave (PAW) pseudopotentials [36] with an energy cutoff of 500 eV. The exchange-correlation energy was described by the generalized gradient approximation (GGA) with Perdew-Burke-Ernzerhof (PBE) parametrization [37]. The Brillouin zone was sampled using the special point scheme of Monkhorst-Pack [38] with a k-point grid of 8×8×8 for the unit cell. In geometry optimization, convergence is achieved when the energy differences are less than $1.0 \times 10^{-4}$ eV, and all the force components must be less than 0.01 eV/Å. Oxygen concentrations were treated with the supercell method, with a 2×2×2 periodicity. The considered range is $0 \leq x \leq 1$.

Since we are treating different oxygen concentrations, the total energy of these structures is not a good parameter to determine the models' stability. Instead, the defect formation energy (DFE) formalism [39], which is independent of the number of atoms, is a good parameter for determining stable systems. In order to apply the DFE formalism, equilibrium thermodynamic is considered between TaN and their constituents, implying that:

$$\mu_{Ta}^{Bulk} + \mu_{N_2}^{Mol} - \Delta H_f = \mu_{TaN} = \mu_{Ta} + \mu_N \qquad (1)$$

where, $\mu_i$ is the chemical potential of the ith species and $\Delta H_f$ is the enthalpy of formation of TaN. Calculated $\Delta H_f$ is -1.70 eV/atom, in agreement with previous reports [40, 41]. Once thermodynamic equilibrium is considered, the DFE takes the following form:

$$E^{DFE} = E^{slab} - E^{ref} - \Delta n_{Ta}\mu_{Ta} - \Delta n_N \mu_N - \Delta n_O \mu_O \qquad (2)$$

Where $E^{slab}$ is the total energy of the system at hand, $E^{ref}$ is the total energy of an arbitrary reference, which in this case is the pristine TaN, and $\Delta n_i$ is the excess or deficit of ith atoms with respect to the reference. In this analysis, all atomic configurations have the same number of Ta atoms, while the N and O atoms vary. Therefore, we vary the chemical potential to evaluate the DFE in a range from O-rich conditions with $\mu_O = \mu_{O_2}^{Mol}$ to O-poor conditions with $\mu_O = \mu_{O_2}^{Mol} - E^{cohesive}$, where $E^{cohesive}$ is the cohesive energy for the oxygen atoms.

## 3. Results and discussion

### 3.1 Chemical composition

The chemical composition of the thin film was identified from the XPS survey spectrum shown in **Figure 1(a)**, in which its binding energy ranges from 0 to 1400 eV; the appearance of several characteristic peaks such as Ta 4f, N 1s, and O 1s reveals that the sample is composed of tantalum, nitrogen, and oxygen only. The C 1s peak is not found, even though carbon is a common contaminant. The inset in **Figure 1(a)** also shows the high-resolution spectrum of O 1s from 525 to 535 eV, where its maximum binding energy appears at 531.2 eV. The peak area was obtained in the CasaXPS program after the background noise subtraction. It is worth mentioning that we did not intentionally put oxygen into the system. Therefore, oxygen is a native impurity in the sample, and it is located on the surface and throughout the material since the sample was isolated to the environment before and during analysis. Oxygen appeared even though the film was deposited with a controlled pressure of ultra-high purity $N_2$.

**Figure 1(b)** shows the high-resolution spectrum of the region containing the Ta $4p_{3/2}$ and N 1s peaks from 390 eV to 412 eV where the maximum binding energy appears at 397.9 eV; it also shows the curve fitting of each peak separately, in which the N 1s associated to tantalum nitride was proposed at 397.7 eV, and the N 1s associated to tantalum oxynitride was proposed at 398.5 eV; on the other hand Ta $4p_{3/2}$ was proposed at 402.9 eV. The envelope exhibits a good fit with the experimental data. The black line in the graph corresponds to the experimental data, the blue line to the N 1s associated to Ta-N, the green line to the N 1s associated to Ta-ON, the dark yellow line to the Ta $4p_{3/2}$, the cyan line to the background, and the magenta line to the fitting envelope. The shift to higher binding energies of the Ta $4p_{3/2}$ signal confirms the Ta-N bond formation since the signal for metallic Ta is reported at 400.9 eV [42] and for TaN at 403.0 eV [43].

**Figure 1(c)** shows the high-resolution spectrum of the Ta 4f peak from 19 to 32 eV. The spectrum exhibits the characteristic doublet constituted of the Ta $4f_{7/2}$ and Ta $4f_{5/2}$ peaks with a separation of 1.9 eV, which coincides with the values reported in the literature [44]. Ta $4f_{7/2}$ appears at a binding energy of 24.0 eV and Ta $4f_{5/2}$ at 25.9 eV. The Ta 4f peak is fitted with three doublets: tantalum nitride at 24.0 eV, tantalum oxynitride at 24.9 eV, and tantalum oxide at 26.3 eV; those components were set with the same doublet separation of 1.9 eV and an area ratio of 4:3 for Ta $4f_{7/2}$ and Ta $4f_{5/2}$. The black line in the graph corresponds to the experimental data, the blue lines to the tantalum nitride, the green lines to the tantalum oxynitride, the red lines to the tantalum oxide, the cyan line to the background, and the magenta line to the fitting envelope. These results suggest that the tantalum can be bonded with nitrogen, oxygen, or both simultaneously. Unfortunately, no depth profile was made. In contrast, an example of how the Ta 4f peak appears when tantalum nitride has oxygen impurities at the surface region can be seen in the work of Nieto *et al.* [45] since there it is not possible to appreciate the Ta doublet in the peak of Ta $4f_{7/2}$ and Ta $4f_{5/2}$, but only after a depth profile of 30 and 90 s.

**Figure 1(d)** shows the valence-band spectrum from -4 to 3 eV; there are unoccupied states at 0 eV; therefore, the thin film is metallic. In addition, the maximum intensity of the binding energy appears at 0.9 eV.

The quantitative analysis of the Ta 4f, N 1s, and O 1s high-resolution peaks reveals that the elemental composition of the thin film in atomic percentage corresponds to 49%Ta, 25%N, and 26%O, which can be represented approximately by the formula $TaO_{0.5}N_{0.5}$ to show that oxygen atoms may replace the nitrogen atoms.

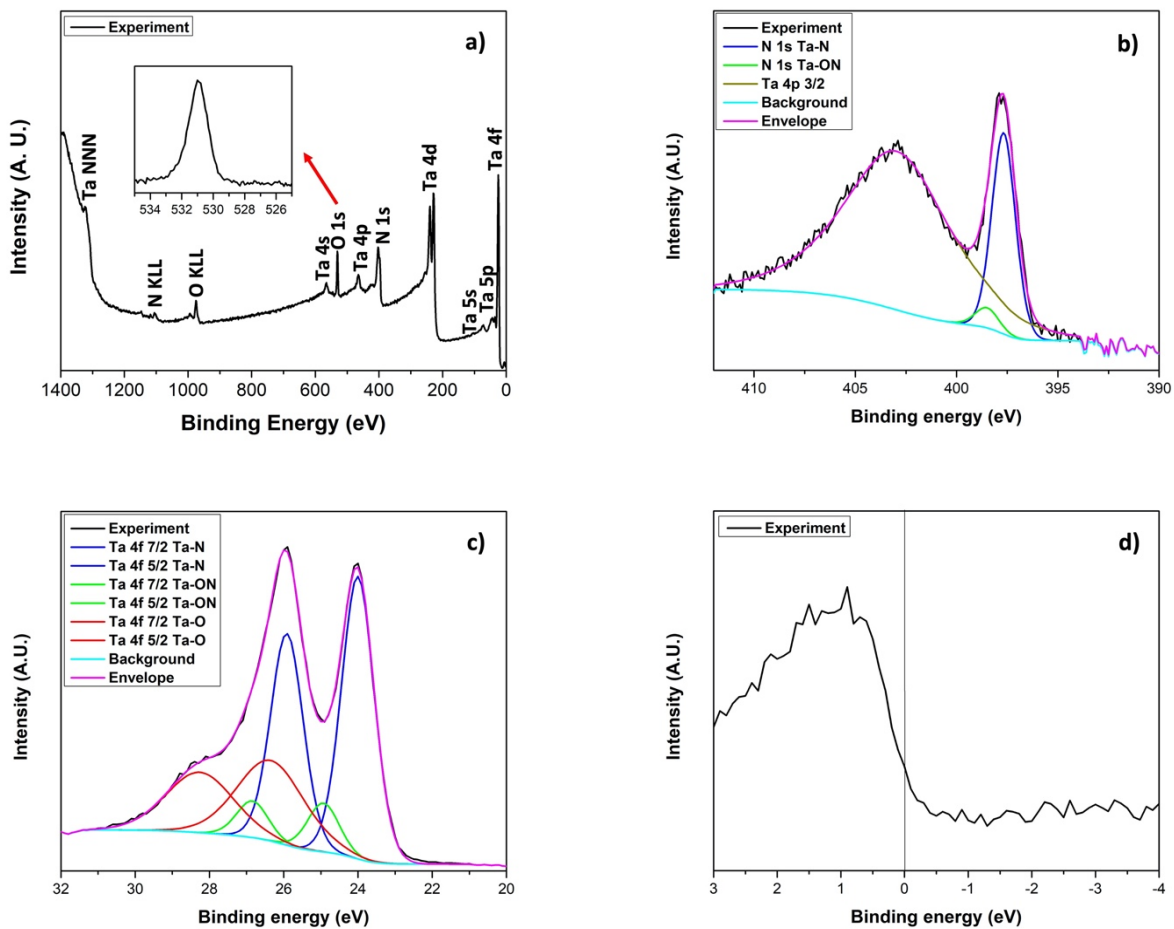

**Figure 1**. Chemical analysis of the thin film by XPS. a) Survey low-resolution spectrum and O 1s high-resolution spectrum, b) N 1s and Ta $4p_{3/2}$ high-resolution spectrum, c) Ta 4f high-resolution spectrum, and d) Valence band high-resolution spectrum.

## 3.2 Crystal structure

**Figure 2(a)** depicts the XRD pattern, plotted in a range of 2θ from 30° to 80°, for the TaN sample. Several peaks appear in the diffractogram, revealing a polycrystalline film. The peaks at 42.850º and 62.290º correspond to the (200) and (220) planes from the MgO substrate, as confirmed in the 00-004-0829 crystallographic card. The remaining peaks, related to the thin film (111), (200), (220), and (311) planes, are indexed using the 00-049-1283 (FCC TaN) card and the 03-065-6750 (FCC TaO) card, which present peaks at 35.831, 41.606, 60.258, and 72.125° for TaN, and at 35.470, 41.187, 59.660, and 71.364° for TaO, respectively. In the experiment, two peaks appear very intense at 41.845 and 60.415°, attributed to the (200) and (220) reflections of the TaN phase. The less intense peak at 35.415º is attributed to the (111) reflection of both TaO and TaN; however, it is closer to TaO; the second peak at 72.415° is related to the (311) reflection of TaN. The thin film lattice parameter calculated from the most intense peak is 4.32 Å.

The values obtained in the experiment are similar to those reported by the previously mentioned ICDD cards, so the presence of the δ-$TaN_{1-x}$ and γ-TaO phases in the thin film is verified. Obtaining the δ-$TaN_{1-x}$ phase by PLD is possible because when the nitride is formed, the Ta-N system is in a state of high excitation produced by the laser energy. The relaxation of the Ta-N system occurs in an extremely short time that does not allow the atoms to rearrange themselves into another structure. So, the desired crystalline phase is obtained even though it is metastable. The γ-TaO appears due to the residual O impurities inside the chamber. It is worth mentioning that although the $Ta_2O_5$ phase is the most stable and known for tantalum oxide, the γ-TaO shares the same space group $Fm\bar{3}m$ with δ-$TaN_{1-x}$ and has a reported lattice parameter of 4.44 Å [46, 47].

The TEM microscopy is used to characterize the film's local crystalline growth, as seen in **Figures 2(b)-2(d)**. The micrographs and diffraction patterns of the thin film cross-section contain information about the local crystal structure of the material and its interplanar distance. **Figure 2(b)** displays a micrograph recorded at low magnification of the cross-section of the sample prepared by FIB for the TEM, showing the MgO substrate, the 35 nm TaO$_x$N$_{1-x}$ thin film, and the Au and C layers used to protect the sample from the FIB.

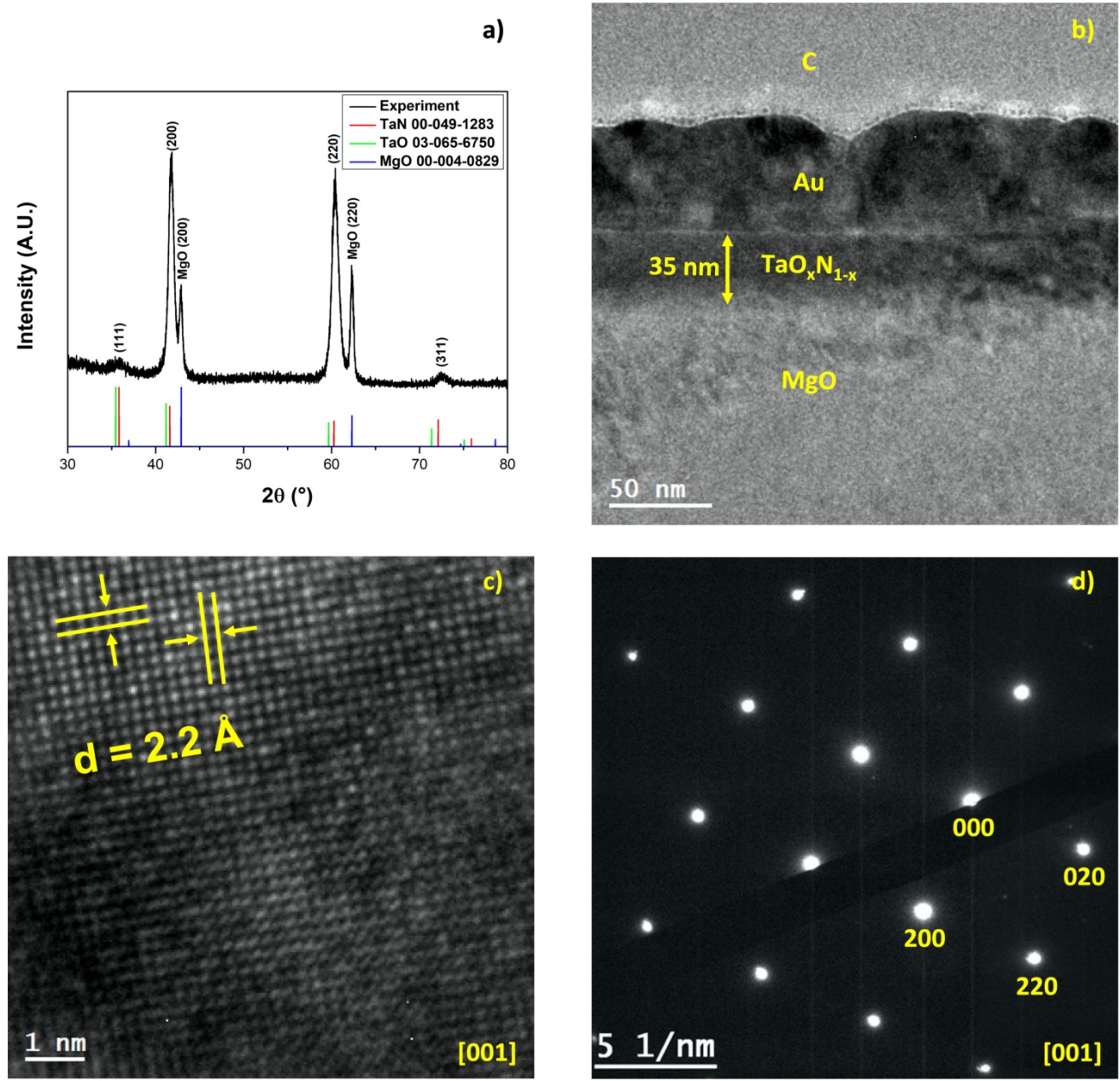

**Figure 2**. Crystal structure of the thin film by XRD and TEM. a) X-ray diffraction pattern, b) Low-magnification micrograph, c) High-magnification micrograph, and d) Electron diffraction pattern.

A high-resolution micrograph of the TaO$_x$N$_{1-x}$ thin film cross-section appears in **Figure 2(c)**; the micrograph is located in a region far from the substrate, near the edge of the film. The local growth remains in the substrate direction and presents the FCC phase with an interplanar distance of 2.2 Å, associated with the (200) and (020) planes. **Figure 2(d)** shows the thin film's indexed selected area electron diffraction (SAED) pattern. An FCC structure is confirmed, and the (200), (020), and (220) planes are associated with it.

From the X-ray diffractogram, the FCC crystalline phase of TaN is detected, as seen in **Figure 2(a)**. Also, it is noticed one peak related to the (111) plane of FCC TaO. No diffractions of monoclinic TaON or orthorhombic $Ta_2O_5$ phases appear. That demonstrates the control in our experiment since the phases mentioned above are the most stable. These phases are absent in the micrographs as well. In low-magnification micrographs, it is observed that the film thickness is ~35 nm. Since our sample was grown at 850 ºC, the obtained preferential orientation is the (200) reflection as we shown in the XRD pattern, which coincides with the behavior reported by Elangovan *et al.*, which evidenced that above 600 ºC, the preferential orientation changes from (111) to (200) [48]. The FCC structure is confirmed based on the TEM micrographs and electron diffraction results. Our results align with the ones by Cristea *et al.*, which report coexisting δ-$TaN_{1-x}$ and γ-TaO in FCC tantalum oxynitride ($TaO_xN_y$) thin films [23].

### 3.3 Superconductivity measurements

**Figure 3** shows the resistance vs. temperature curve of the tantalum oxynitride thin film; the area of interest is plotted from 0 to 20 K. The film presents superconductivity at a $T_c$ of 2.99 K, because its resistance dropped abruptly in a transition from 306.11 Ω (at 5.43 K) to 0 Ω (at 2.99K). Although magnetic susceptibility was not measured, the abrupt drop in the curve allows us to identify that the sample presents the zero-resistance characteristic at 2.99 K; to confirm that value, the instrument measured 1413 points in that zone. Some authors report the $T_c$ at the onset of the transition [12]. In our case, the onset is located at 5.43 K, the midpoint of the drop at 4.30 K, and the zero-resistance zone at 2.99 K, so the transition has a width of 2.44 K. The superconductivity of the tantalum oxynitride film is a consequence of the presence of the FCC structure, as previously discussed in the literature for tantalum nitride [4]. The stoichiometry has less weight than the crystalline structure in the superconductivity since the film does not lose its superconducting character, even when a considerable amount of oxygen impurities (26 %O) appear. Also, superconductivity was found recently in bulk FCC TaO with $T_c$=6.20 K [46], so it is expected that the possible TaO regions formed in the thin film are superconducting as well. The consequence of having random oxygen impurities in the sample is that the film $T_c$ is lower than the one reported by Reichelt *et al.*, where they found a $T_c$=10.8 K [7]. However, our obtained $T_c$ is within the range of the reported values for TaN thin films by other authors as ~4.31 - 8.15 K [11], 5.1 - 10.8 K [7], 5.9 - 7 K [12], 6 - 8 K [14], and 0.5 - 6.4 K [13]; or with the bulk TaN $T_c$=6.5 K [15].

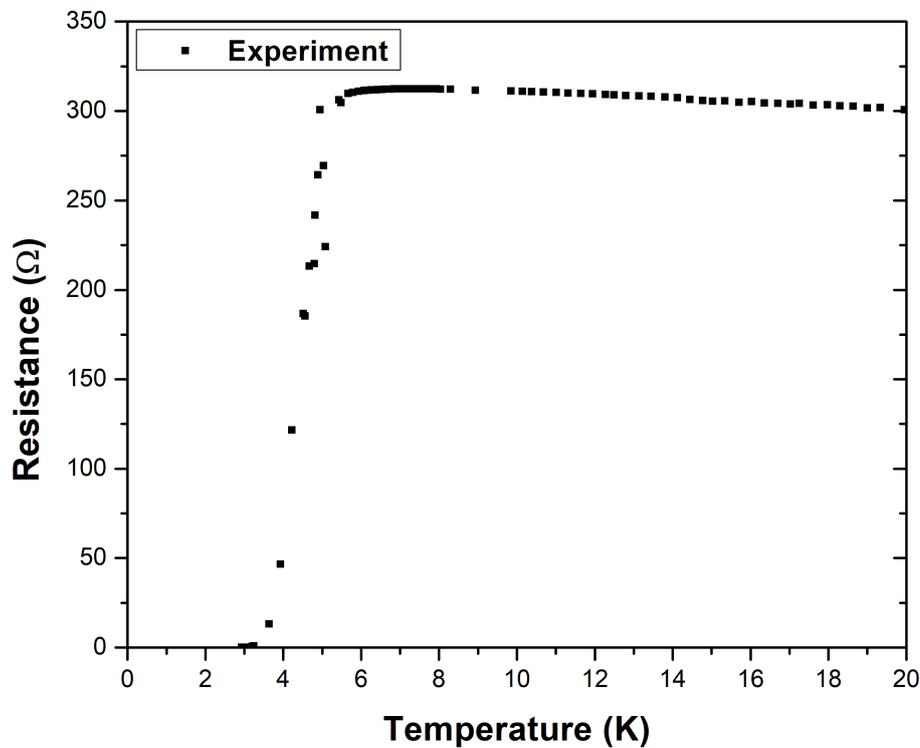

**Figure 3**. Resistance vs. Temperature curve of the thin film measured by the four-point probe method.

In other words, the oxygen contained in the sample prevents the tantalum nitride films from reaching their maximum possible $T_c$. Notice that even at highly controlled growth conditions, oxygen impurities appear. The differences in the reported $T_c$ indicate that the role of oxygen impurities -in the superconductivity performance- still needs to be discussed appropriately in the literature. Also, it is evident that TaN is highly reactive to oxygen, and even at ultra-high vacuum, the remnant oxygen - maybe accompanying the N flow- would be present in the thin films. To corroborate our assumptions, we simulate -at an atomic scale- the oxygen capacity to enter into the FCC TaN lattice.

**3.4 Atomic-scale analysis of the oxygen incorporation into TaN**

In this section, by DFT calculations, we investigate the presence of oxygen in the TaN structure. The calculations use TaN as a reference. The highest O content is modeled as the TaO structure. TaN and TaO are isostructural. The calculated lattice parameter for TaN is 4.42 Å with a Ta-N bond distance of 2.21 Å, and the cell parameter for TaO is 4.51 Å with a Ta-O distance of 2.25 Å, in agreement with previous experimental reports [8, 46]. To study different oxide content, a 2×2×2 periodicity is employed. The system contains 64 atoms, 32 Ta and 32 that may be N, O, or both.

**Table 1.** Oxygen content in the $TaO_xN_{1-x}$ supercells.

| System | Number of O atoms | %O | Lattice parameter (Å) |
| --- | --- | --- | --- |
| TaN | 0 | 0.00 | 8.84 |
| $TaO_{0.03}N_{0.97}$ | 1 | 1.56 | 8.85 |
| $TaO_{0.06}N_{0.94}$ | 2 | 3.13 | 8.85 |
| $TaO_{0.13}N_{0.87}$ | 4 | 6.25 | 8.86 |
| $TaO_{0.19}N_{0.81}$ | 6 | 9.38 | 8.90 |
| $TaO_{0.25}N_{0.75}$ | 8 | 12.50 | 8.93 |
| $TaO_{0.50}N_{0.50}$ | 16 | 25.00 | 9.03 |
| TaO | 32 | 50.00 | 9.01 |

**Table 1** summarizes the cell parameters for each considered stoichiometry. It is noticed that as the oxygen amount increases, the cell parameter expands up to 50% oxygen content. After that, the cell parameter contracts until it reaches the TaO lattice parameter. The models for each oxygen content are in the supplementary section. We only show the TaN, TaO$_{0.5}$N$_{0.5}$, and TaO structures (see **Figure 5**). Oxygen is randomly distributed to avoid initial clustering. In the experiment, few oxygen atoms were available in the ultra-high vacuum chamber. As discussed in the experimental section, insufficient oxygen is available to get clustering. Also, experimental evidence suggests that oxygen is taking on the places of N atoms in the TaN lattice.

**3.5 Thermodynamic stability of the tantalum oxynitrides**

To analyze the viability oxygen has to replace nitrogen atoms from the TaN lattice, we proceed to evaluate the defect formation energy. **Figure 4** depicts the formation energy vs. oxygen chemical potential from O-poor to O-rich conditions. The energy reference is the bare TaN structure. Stable models must have lower energy than the reference (negative defect formation energy); otherwise, they are unstable. In **Figure 4**, the left stands for O-poor conditions, while the right stands for O-rich conditions. Notice that the bare TaN is the most favorable model at O-poor and intermediated conditions, an expected fact since there is a lack of oxygen under O-poor conditions.

On the other hand, all substitution models are viable at O-rich conditions -such energies demonstrate the spontaneous oxygen incorporation-, even the TaO$_{0.03}$N$_{0.97}$ system, the one with the lowest oxygen content. The stability of each model increases as oxygen atoms replace nitrogen from the TaN lattice until they reach the TaO$_{0.50}$N$_{0.50}$ system. After that, it is expected to decrease in stability until reaching TaO. Notice that we included TaO as a limit case. However, in the experiment, it is not expected to observe such oxygen quantities since we are working at an ultra-high vacuum system. As mentioned in the experimental part, the small oxygen content in the chamber may be due to the N$_2$ flux. It should be noted that the compositions studied in this work are less stable than hexagonal TaN, monoclinic TaON, and orthorhombic Ta$_2$O$_5$ (the most stable phases of the three compounds). Still, the FCC TaN can be induced using a Physical Vapor Deposition technique on MgO, Si, or sapphire substrates [1, 7, 12, 14].

Our defect formation energy analysis evidences that even at low oxygen concentrations, the TaN is proclive to oxidize, so tantalum oxynitride must always be considered when discussing the superconductivity measurements, which has been overlooked in the existing literature.

As another piece of evidence about the spontaneous formation of tantalum oxynitride, some authors have reported that ε-TaN may not exist as a binary nitride [49]. Still, it stabilizes with oxygen atoms, a hypothesis that matches our results for δ-TaN$_{1-x}$.

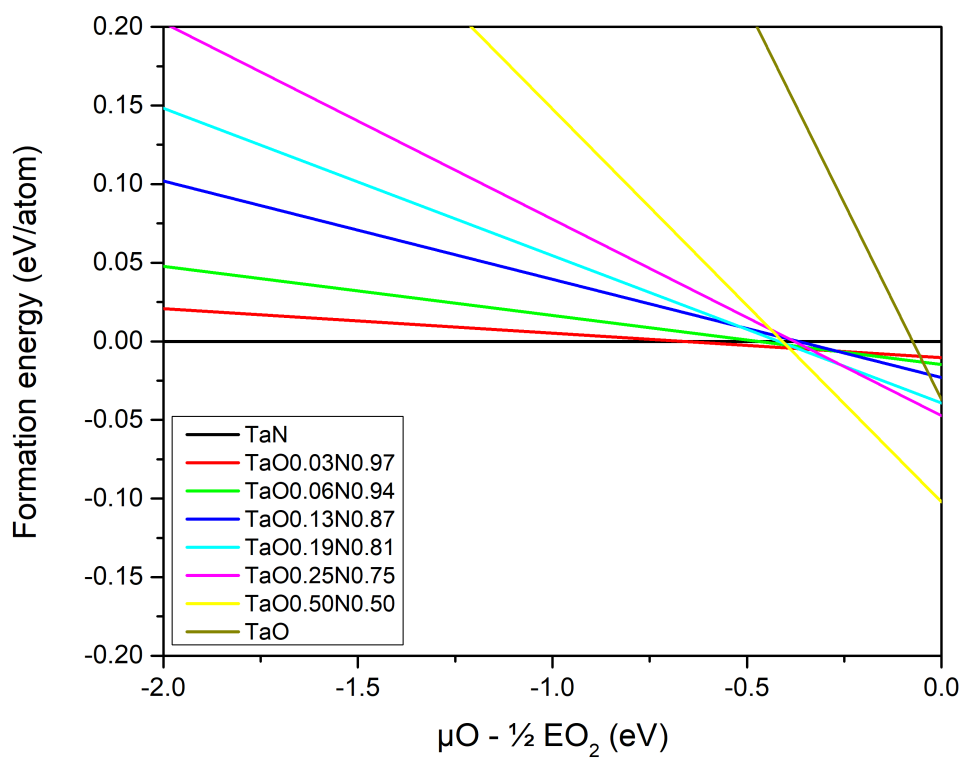

**Figure 4**. Formation Energy vs. Oxygen chemical potential from O-poor to O-rich conditions.

**3.6 Electronic properties**

To analyze the behavior of the most stable tantalum oxynitride structure, we investigated the electronic properties of the TaO$_{0.5}$N$_{0.5}$ model and compare it with the TaN and TaO models by calculating its Density of States (DOS). Fermi level is the energy reference. In all cases, a metallic behavior emerges and agrees with the valence band obtained by XPS measurements, see Figure 1d and Figure 5b. The results for the three systems are depicted in **Figure 5.**

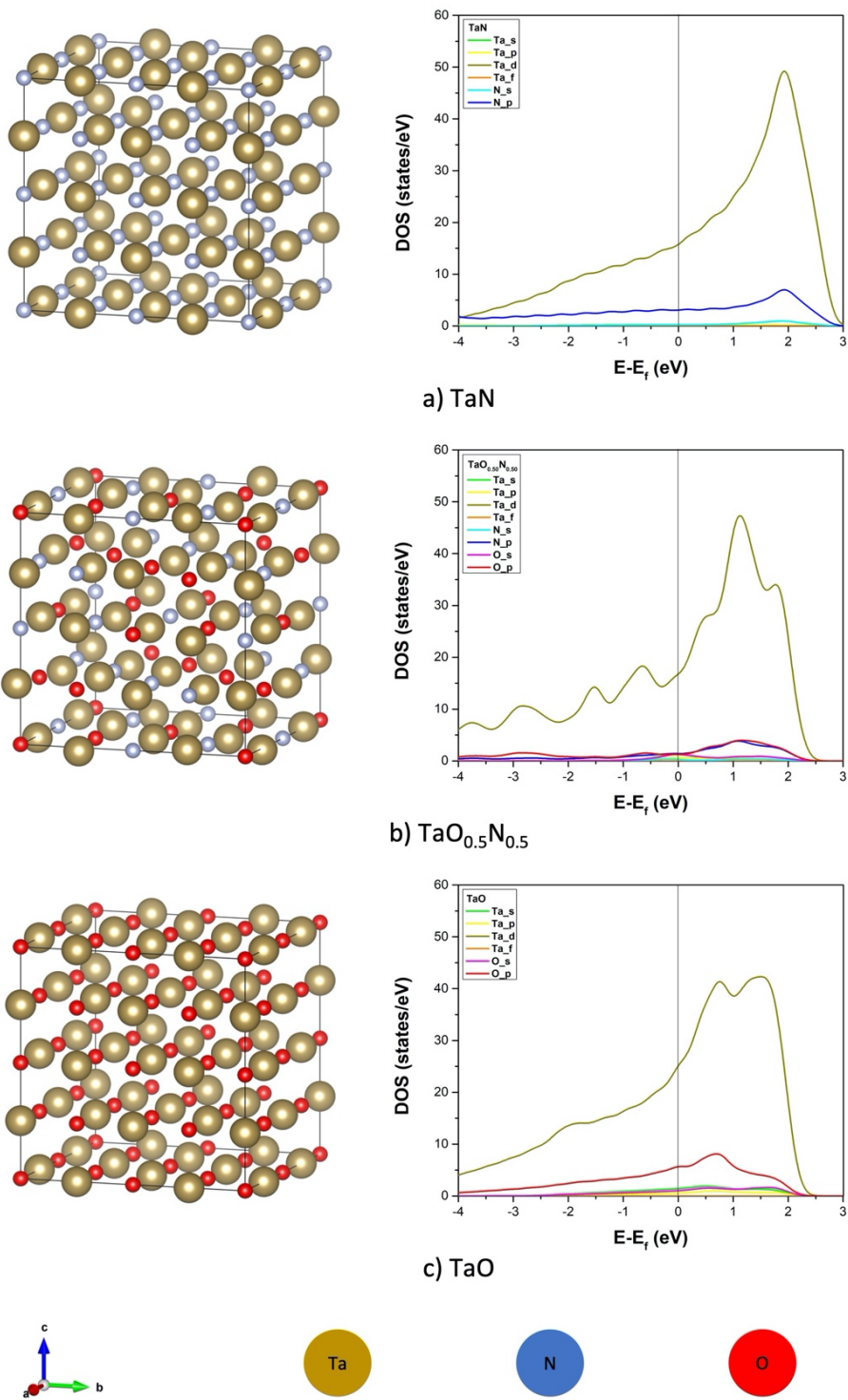

**Figure 5.** Optimized structures and projected Density of States from -4 to 3 eV of (a) TaN, (b) $TaO_{0.50}N_{0.50}$, and (c) TaO. The Fermi level is set to 0 eV. The graphs show the contribution of the *s* and *p* orbitals of N and O and the contribution of *s*, *p*, *d*, and *f* orbitals of Ta.

About the bare TaN, our DOS agrees with the report by Stampfl *et al.* [50], where the main contribution to the Fermi level comes from the Ta-*d* orbitals. A similar behavior is noticed for the $TaO_{0.50}N_{0.50}$ oxynitride, where the Ta-*d* orbitals mainly contribute to the DOS around the Fermi level. In contrast, the N-*p* and O-*p* orbitals contribute almost equally. In the case of tantalum oxide, the main contribution at the Fermi level, again, comes from the Ta-*d* orbitals, followed by the O-*p* contributions.

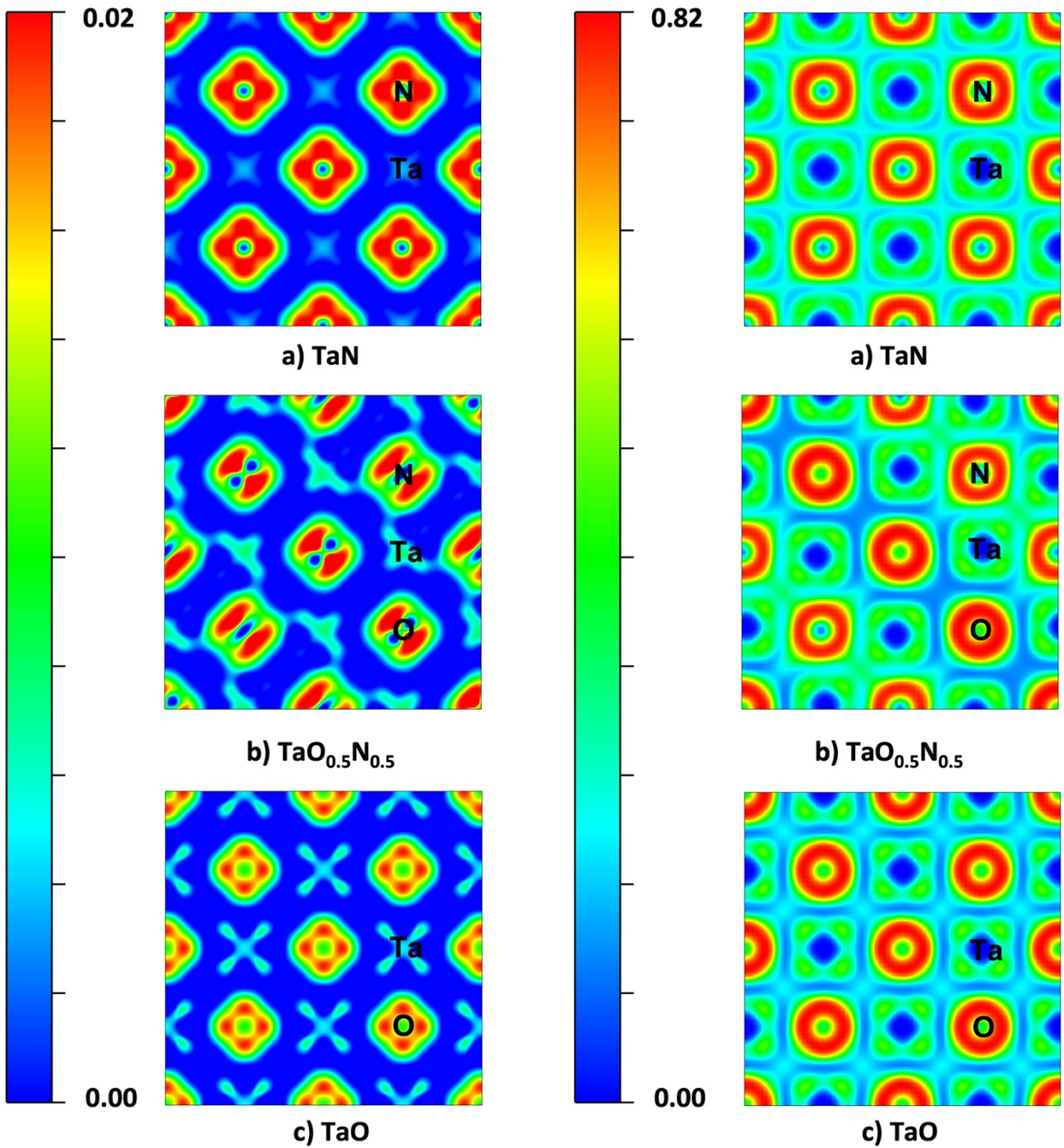

**Figure 6.** A 2D slice view of the CDD from 0 to 0.02 and the ELF from 0 to 0.82 of (a) TaN, (b) TaO$_{0.50}$N$_{0.50}$, and (c) TaO at the (010) plane.

To elucidate the bonding type in the pristine and oxidized systems, we plot the charge density difference (CDD) and the electron localization function (ELF) at the (010) plane. An RGB scheme was used in both cases. The CDD of a) TaN, b) TaO$_{0.5}$N$_{0.5}$, and c) TaO are shown in the left panel of **Figure 6**. Blue is for charge depletion, while red is for charge accumulation. In all cases, the Ta atoms show a depletion of the electronic density. In contrast, N and O atoms denote charge accumulation, suggesting charge transfer between the neighboring atoms and an ionic nature.

To further confirm the ionic character in the structures of interest ELF density maps for the same systems are plotted on the right panel of **Figure 6**. The blue color suggests a low probability of finding electron accumulation; otherwise is characterized by a red color. Tantalum nitride, oxynitride, and oxide show similar behavior, since the main probability to find the electrons is around the electronegative atoms. Also, the non-electron population is noticed in the middle point between Ta-O and Ta-N which suggest an ionic interaction. This fact is corroborated by ELF line profiles shown in **Figure 7**, where the ionic nature of both interactions is easy to observe, however, the Ta-O provides the most ionic interaction since at the middle point of the bond less electron population is observed.

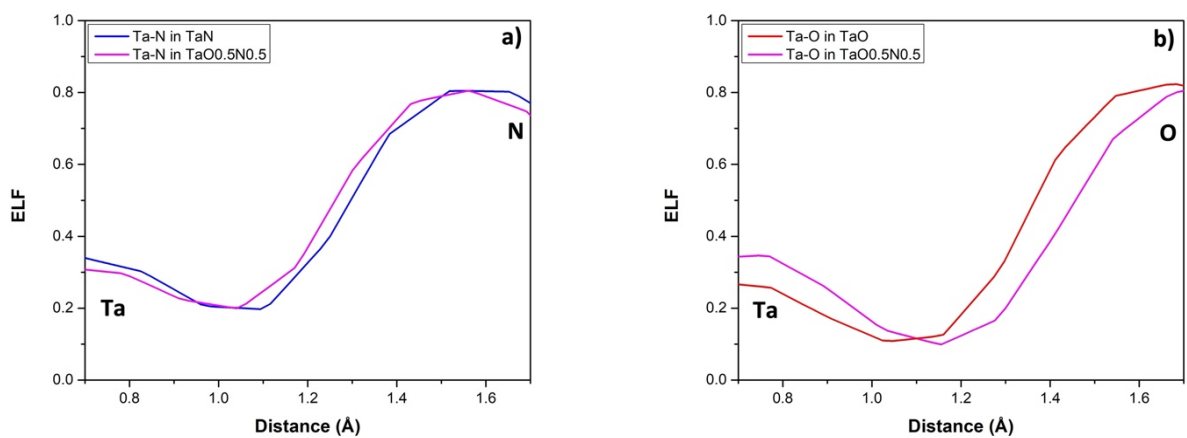

**Figure 7.** Line profiles of the ELF of the (a) Ta-N and (b) Ta-O bonds of the TaN, TaO$_{0.5}$N$_{0.5}$, and TaO supercells.

## 4. Conclusions

Tantalum nitride was synthesized by reactive PLD on a MgO substrate. It showed superconducting characteristics at $T_c$= 2.99 K. XRD analysis confirmed an FCC TaN structure. Not secondary phases were observed. Also, TEM analysis evidenced a local epitaxial growth in the thin film. The lattice parameter of the film is smaller near the substrate and larger as it moves away from it, but in both cases it remains similar. The thickness of the film was 35 nm. The electron diffraction patterns showed that both the film and the substrate present the same crystalline structure (FCC). However, elemental analysis by *in situ* XPS evidences the presence of native oxide in the sample up to 26% at. By DFT calculations and thermodynamics criteria, we have evidence that oxygen can occupy the N sites of the crystal without structural modifications. Our experimental and theoretical findings support the idea that TaN does not exist in pure form but as tantalum oxynitride. Therefore, it is important to consider the presence of oxygen in the superconductor and how it modifies its properties, a fact not considered previously in the literature and may be the reason for the different critical temperatures measured in tantalum nitride materials.

**Acknowledgments**

We thank FORDECYT 272894 project, DGAPA-UNAM projects IG101124, IG101623, IA100624, and IN101523 and CB_CONACYT A1-S-33492 grant for the financial support. Calculations were performed in the DGCTIC-UNAM Supercomputing Center, project no. LANCAD-UNAM-DGTIC-084, LANCAD-UNAM-DGTIC-150, LANCAD-UNAM-DGTIC-368 and LANCAD-UNAM-DGTIC-422. We thank David Domínguez, Eloisa Aparicio, Israel Gradilla, Eduardo Murillo, and Francisco Ruiz for their support in the characterization of the samples. Also, we thank Aldo Rodriguez-Guerrero, J.I. Paez-Ornelas for technical support and useful discussions. C.C.G. Thanks to CONAHCYT for the postdoctoral grant. M.I. Pérez Montfort and A. Cortés Lemus corrected the English version of the manuscript.